# A Framework to Support the Trust Process in News and Social Media

Carlos Laufer[1][+552135271500] and Daniel Schwabe[1] [+552135271500]

[1] Dept. of Informatics, PUC-Rio
R. M. de S. Vicente, 225, Rio de Janeiro, RJ 22453-900, Brazil
`laufer@globo.com, dschwabe@inf.puc-rio.br`

**Abstract.** Current society is heavily influenced by the spread of online information, containing all sorts of claims, commonly found in news stories, tweets, and social media postings. Depending on each user, they may be considered "true" or "false", depending on the agent's trust on the claim. In this paper, we discuss the concept of content trust and trust process, and propose a framework to describe the trust *process*, which can support various possible models of content trust..

**Keywords:** trust, provenance, linked data, social networks, news.

## 1    Introduction

The issue of trust has been present in the Internet shortly after its popularization in the early 90s (see [13] for a survey). The focus has been on lower level of the Internet Architecture, with an emphasis on authentication. More recently, with the advent of the Web and social networks, the cybersphere, and society as a whole, has become heavily influenced by information (and misinformation) that flows in news sites and social networks in the Internet. There are many studies carried out in several disciplines attempting to characterize and understand the spread of information in the cybersphere, and how this affects society (see [18] for an overview). A more visible aspect has been the spread of "fake news", actually a term used to refer to several different misuses of information, as postulated by Wardle in [24]. This has also been the focus of much research and many initiatives (e.g. [8, 9, 3]).

The advent of user-contributed data has raised the issue of "data quality", as evidenced for example in case of online reviews and social media, often with direct effects on commercial success (e.g. [4]). This highlights the fact that data, in reality, expresses a belief or opinion of some agent.

The original vision for the Semantic Web included a "Trust" layer, although its emphasis was more on authentication and validation, and static trust measures for data. There have been many efforts towards representing trust, including computational models - a general survey can be found in [20]; [5] presents an excellent earlier survey for the Semantic Web; and [23] surveys trust in social networks. In the Linked Data world, it is clear that facts on the Semantic Web should be regarded as claims rather than hard facts (e.g., [6]).

Most if not all research surveyed in [20], present what Gerck calls "Trust Models", as opposed to "Models of Trust" [10], i.e. a model of what is the trust process that is being adopted or assumed, where the former presumes the latter. In this paper, we propose a framework to describe the trust *process*, which can incorporate various alternative models of content trust (using the terminology proposed in [11]).

From a more global perspective, the ability to support a trust process over the information available in the cybersphere is directly related to at least two of the UN's Sustainable Development Goals[1] – Peace, Justice and Strong Institutions, and Reduced Inequalities. At the very least, a well-supported trust process can improve Transparency with respect to the actions undertaken by political agents, by allowing members of society to rely on information about these actions to make decisions and eventually take further actions in response.

The remainder of this paper is organized as follows. Section 2 discusses the concept of trust; Section 3 presents a model of the trust process; Section 4 discusses trust regarding information flow in news media and social networks.

## 2 The Concept of Trust

Whereas there are many definitions of trust (e.g. [19]), we base ourselves in the work of Gerck [10] and Castelfranchi et al. [7], taking the view that trust is "knowledge-based reliance on received information", that is, an agent decides to trust (or not) based solely on his knowledge, and the decision to trust implies the decision to rely on the truth of received or known information to perform some action.

Castelfranchi et al. define trust in the context of multi-agent systems, where agents have goals, asserting that trust is "a mental state, a complex attitude of an agent x towards another agent y about the behavior/action relevant for the result (goal) g. This attitude leads the agent x to the decision of relying on y having the behavior/action, in order to achieve the goal g".

Gerck presents a definition of trust as "what an observer knows about an entity and can rely upon to a qualified extent". The two definitions have close parallels: the observer is the agent who trusts; the entity is the trusted agent; the qualified extent is the behavior/action. Both associate trust with reliance. However, the former definition mentions explicitly the goal-oriented nature of trust.

From both definitions, we observe that trust implies reliance: when an agent trusts something, it relies on its truth to achieve some goal without further analysis – even if it is running the risk of taking an inappropriate or even damaging action if the object of trust is false.

## 3 A Framework for the Trust Process

Given the considerations above, we present a model for the trust process that underlies the use or consumption of data/information on the web, represented diagrammatically in Figure 1.[2]

We focus here on the cases where an *Agent* needs to act, i.e., do some computation, to decide, or take some *Action*. The agent must act based on some *Data/Information* items which, clearly, it must trust – the *Trusted Data*. The *Data/Information* items to be used by the agent may come from several sources, and it is not always clear (to the *Agent*) what is the quality of this *Data*, or the trustworthiness of the *Information* it contains. Therefore, the *Agent* must apply a *Trust Process* to filter the incoming *Data/Information* items and extract the *Trusted Data* items to be used by the *Action*. From this point of view, an item is considered as the smallest indivisible element that can be used in the *Trust Process* and may have an internal structure when used by the *Action*.

---

This *Trust Process* can be based on a multitude of different signals, some of which we have singled out in the diagram in Figure 1, namely, the *Metadata* which describes various properties of the *Data/Information* items, and the *Context* in which the *Action* will take place. The criteria used in the *Trust Process* are expressed by *Policies* determined by the *Agent*. Notice that in this framework, the *Context* contains any arbitrary information items used by the *Policies*, in addition to *Metadata* and the *Data/Information* items themselves.

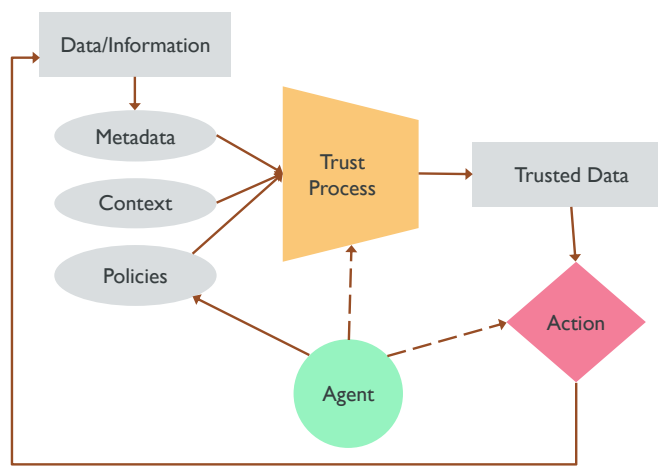

**Figure 1** – A Framework of the Trust Process. Continuous lines represent "consume/produce"; dashed lines represent "executes".

Many, if not most, of the trust models discussed in the surveys mentioned in Section 1 [5, 23] can be regarded as providing models or representations for one or more of the elements of this trust framework, such as different representations for the metadata, or specification language for the policies, or particular types of context information that can be added to the data. Regardless of these models, it should be clear that, as far as the *Action* is concerned, the whole process is *binary*: An incoming *Data/Information* item is either accepted and added to the *Trusted Data*, or it is not – there are no "half filters". In other words, when time comes to *Act*, either the *Agent* uses that *Data/Information* item, or it does not, it cannot "use it partially" – an element is considered indivisible from the point of view of the *Trust Process* [2].

Thus, the actual trusting process ("Model of Trust") should not be confused with models which may attribute non-discrete or continuous values to the trustworthiness assigned to data/information items that are used to determine if the incoming data/information items should be filtered or not ("Trust Models").

To further illustrate this difference, the earlier approaches such as [12, 21] propose to compute measures of trust by treating the RDF graph of the statements about the relations between resources a social network graph. The trust process is implicit in the subset of RDF properties (relations) considered. For instance, in [12], for trust between persons, an additional relation indicating the trust level of one person on another person, for a given "subject" (topic). Given one such graph, it is possible to extract a sub-graph with respect to a subject of interest, and then compute the proposed measures over this graph. This model does not explicitly deal with the notion of claims and how the "trusted graph" is actually built. Furthermore, it does not discuss how trust policies, including automated ones, could be integrated, and does not directly identify who the trusting agent is.

The work reported in [6] introduces the notion of information filtering, in the broader context of information quality, one of which is trustworthiness. In this respect, it is similar to the general schema proposed here, but does not directly identify the specific components to enable trust, as outlined here. Within this view of a filtering process, trust policies can be expressed in suitable languages, such as the ones proposed in [2, 15], which would be consistent with our proposed approach, but requiring extensions to represent the elements of the model proposed here, such as the trusting agent and the action to be performed.

Another point to notice is that, in this framework, only the *Agent* determines the policies it wishes to apply to incoming *Data/Information*. The metadata associated with the incoming Data/Information may contain *Data/Information* about the publisher or provider of a *Data/Information* item, and the *Policies* may take this into account in the filtering process.

In contrast, privacy issues would require adding a similar set of metadata and policies to the *published* Data/Information that will be consumed by the *Agent*, which would act as an additional filter, applied before the Data/Information is made available to the *Agent*. The privacy *Policies* of a Publisher may use information about the (requesting) *Agent*, the *Action* and the *Context*.

In this work, we will focus on specific types of *Data/Information* – those pertaining to Political Systems, and one specific type of metadata, *Provenance*, in the context of news media. In the next section, we discuss the trust process in this context, and propose a model for information and metadata to support it.

## 4 News Media, Social Networks and Provenance

The nature of information in news and social media must, necessarily, be seen as containing claims made by someone about some "fact" (or "statement"), i.e., someone (an agent, i.e. a person or organization) claims a certain fact to be true. It is up to the reader to accept the truth of these claimed facts.

For an agent $A$ to accept the truth of a certain claim about a hitherto unknown fact $F$ entails either establishing the truth of $F$ through direct observation[3], (or a chain of interdependent observations), or trusting the source $S$ of the claim. The trust towards $S$ then implies that one expects the claim to be true because $S$ has been consistently (predictably and expectedly) correct in the past.

If, however, $S$ is not trusted, one may require some evidence to serve as the basis to establish the truth of the claim. Such evidence is, in turn, another (possibly a set of) fact(s) $F1$ in a claim made by some third-party $SS$, which $S$ believes is more trusted by A than itself ($S$). In this fashion, a chain of interdependent claims is established, such that the truth of $F$ relies on the trust towards the agent at the end of the chain ($SS$ in this example).

In most societies, certain public agents (e.g., Vital Record Office, Notary Public) have, by convention and mutual agreement among its members, the so-called "public faith", meaning that whatever they claim (about specific kinds of facts) is deemed "true". For instance, Obama's nationality was determined by a birth certificate which "proved" as far as the law is concerned that he is indeed a US citizen.

From the discussion above, it is safe to say that provenance data about information published in the cybersphere is crucial to enable agents to consume this information – i.e., accept it as true and eventually act based on this. This has been clearly recognized by many researchers and scholars studying the "fake news" phenomenon, at least with respect to "fabricated content" as outlined in [24].

We have been involved with an initiative to publish a database about Political Agents in Brazil in the form of Linked Data, named "Se Liga na Politica"[4] (SLNP). The data in this database is obtained from several sources, in both automated and non-automated ways. Most of the automated extraction is made from official sources, such as the open data published by the House of Deputies and by the Senate. In addition to such sources, data may also be contributed by individuals, in crowdsourced fashion. One of the main usages for this database is to provide context information for news stories, to allow readers to establish trust in the claimed facts based on their own criteria. We present here the rationale and design decisions for the underlying ontologies used in this database, starting first with our motivations.

---

[3] It is of common sense that a person normally trusts her/himself as a source of information, although s/he may defer on this perception if someone s/he believes to be wiser (with respect to this particular subject matter) contradicts her/him. Furthermore, current technology can successfully trick human perception. (e.g. Google Duplex - https://goo.gl/mG4E9h).

We regard a Political System as sets of agents and organizations that govern a society. Our goal in the SNLP project is to help make the existing relationships between political agents in a Political System explicit and thus subject to analysis. We do so by providing a Linked Data database where such relationships appear as data items. Given the multiplicity of sources, and the nature of the subject matter, this database is designed so that facts are seen as claims made by some agent, and therefore provenance information becomes a "first class citizen" of the domain.

The SLNP database represents claims based on the nanopublication model [14], where a claim, represented by a set of statements, is recorded as a group of three named graphs:

- assertion graph - containing the set of statements that compose the claim, using the POLARE family of ontologies [17];
- publication info - containing statements that give provenance information about the claim publication activity, using the provHeart PROV-O pattern [16];
- provenance - containing statements that give provenance information about the claim itself, using the provHeart pattern.

Further details on how Provenance can be used to support the trust process can be found in [16].

## 5    Conclusions

In this paper we have presented and explicit model of the trust process which can be seen as an evolution of a collection of previous work by several authors. It makes explicit the elements of the trust process – Trusting Agent, Action, Information, Meta-Information, Context and Policies. We have also outlined how this model is well suited to support the use of news and social media information, by representing this information as claims and recording provenance.

Ongoing and future work will examine suitable policy languages; how to mine implicit policies employed by humans when consuming online information; how appropriate narratives can be constructed and to better convey trusted information to human beings, and how this can be effectively communicated in human-computer interfaces.